# Brightened spin-triplet interlayer excitons and optical selection rules in van der Waals heterobilayers

Hongyi Yu[1], Gui-Bin Liu[2], Wang Yao[1*]

[1]Department of Physics and Center of Theoretical and Computational Physics, University of Hong Kong, Hong Kong, China
[2]School of Physics, Beijing Institute of Technology, Beijing 100081, China
[*]Correspondence to: wangyao@hku.hk

**Abstract:** We investigate the optical properties of spin-triplet interlayer excitons in heterobilayer transition metal dichalcogenides in comparison with the spin-singlet ones. Surprisingly, the optical transition dipole of the spin-triplet exciton is found to be in the same order of magnitude to that of the spin-singlet exciton, in sharp contrast to the monolayer excitons where the spin triplet species is considered as dark compared to the singlet. Unlike the monolayer excitons whose spin-conserved (spin-flip) transition dipole can only couple to light of in-plane (out-of-plane) polarisation, such restriction is removed for the interlayer excitons due to the breaking of the out-of-plane mirror symmetry. We find that as the interlayer atomic registry changes, the optical transition dipole of interlayer exciton crosses between in-plane ones of opposite circular polarisation and the out-of-plane one for both the spin-triplet and spin-singlet species. As a result, excitons of both species have non-negligible coupling into photon modes of both in-plane and out-of-plane propagations, another sharp difference from the monolayers where the exciton couples predominantly into the out-of-plane propagation channel. At given atomic registry, the spin-triplet and spin-singlet excitons have distinct valley polarisation selection rules, allowing the selective optical addressing of both the valley configuration and the spin singlet/triplet configuration of interlayer excitons.

## I.  Introduction

Monolayer transition metal dichalcogenides (TMDs) have recently emerged as a new class of direct-gap semiconductors in the two-dimensional (2D) limit [1-4]. The visible range direct bandgap located at the hexagonal Brillouin zone (BZ) corners, called $\pm \mathbf{K}$ valleys, makes these monolayer semiconductors ideal platforms for optoelectronics, with a possibility to address the valley pseudospin through the valley contrasted polarisation selection rules [5-10]. The atomically thin geometry of the monolayer greatly enhances the Coulomb interaction, thus the optical response of monolayer TMDs is dominated by the Wannier type exciton [11], the hydrogen-like bound state of an electron-hole pair formed at $\pm\mathbf{K}$ valley. Excitons in monolayer TMDs are found to exhibit exceptionally large binding energies of hundreds of meV [12-16], small Bohr radius of $\sim$ nm [17], and strong coupling to the light [18,19].

The strong spin-orbit-coupling (SOC) of the transition metal atom introduces a spin splitting of several hundred meV at the valence band $\pm\mathbf{K}$ valleys [9]. This leads to the spin-valley locking of the band edge holes, i.e., a low energy hole in $\mathbf{K}$ ($-\mathbf{K}$) valley corresponds to spin-down (-up). On the other hand, the conduction band spin splitting at $\pm\mathbf{K}$ valleys is also finite ($\sim$30 meV in MoSe$_2$, WSe$_2$ and WS$_2$ [20-22]). The electron and hole constituents of an exciton can either have antiparallel spins (referred hereafter as spin-singlet exciton) or parallel spins (referred hereafter as spin-triplet exciton). The conduction band spin splitting therefore contributes to the energy splitting of the spin singlet and triplet excitons. With the spin-down (-up) hole defined as the vacancy in the spin-up (-down) valence band, the spin-singlet exciton is created/annihilated through the spin-conserving interband optical transition, while the spin-triplet exciton is created/annihilated through the spin-flip interband transition.

The valley-dependent polarisation selection rules for the excitons in monolayer TMDs have been well studied [5-9,23]. Earlier works focused on the spin-singlet excitons, while the spin-triplet ones were considered as optically dark. Due to the $2\pi/3$-rotational ($\hat{C}_3$) symmetry and the out-of-plane mirror ($\hat{\sigma}_h$) symmetry of the monolayer lattice structure [2], a spin-singlet exciton at **K** ($-$**K**) valley has a in-plane transition dipole, which couples to a $\sigma+$ ($\sigma-$) circularly polarised photon propagating in the out-of-plane ($z$) direction [9]. However, it is found recently that the band edge spin-triplet exciton in monolayer WSe$_2$ is not completely dark, but has a very weak transition dipole pointing in the $z$-direction [23-26]. The radiative recombination of such a spin-triplet exciton, which corresponds to a spin-flip interband transition, albeit weak, can emit a $z$-polarised photon propagating in the in-plane direction [23-26].

Two different TMDs monolayers can be vertically stacked to form a van der Waals heterobilayer, which provides an approach to engineer layered heterostructures for novel optoelectronic devices [27]. The heterobilayer has a type-II band alignment, which means the conduction and valence band edges are located in different layers. Excitons in such heterobilayer then have the interlayer configuration as the ground state, which dominates the low energy optical properties of the heterobilayers [28-32]. With the small interlayer separation, such interlayer excitons are expected to have Bohr radius and binding energy in the same order of magnitude to those of monolayer excitons [33].

The interlayer exciton luminescence properties are very different from the monolayer case due to the following facts: (1) The electron-hole vertical separation substantially reduces the interlayer exciton oscillator strength, which is found to be two to three orders of magnitude weaker than that of the monolayer exciton [34]. (2) The photoluminescence (PL) from the interlayer excitons can only be observed when the two layers have nearly aligned crystalline directions [28]. In such heterobilayers a large scale moiré superlattice pattern forms [35], where the exciton properties vary from local to local [36,37].

In this paper, we investigate the optical properties of spin-triplet interlayer excitons in heterobilayer TMDs in comparison with the spin-singlet ones. We show that, unlike the monolayer excitons whose spin-conserved (spin-flip) interband transitions can only couple to photons with in-plane (out-of-plane) polarisation, such restriction is removed for the interlayer excitons due to the breaking of the out-of-plane mirror symmetry in the heterobilayers. In a lattice-matched heterobilayer, we establish a one-to-one correspondence between the interlayer atomic registry and the optical transition dipole. For both the spin-singlet and spin-triplet interlayer excitons, the optical dipole can point either in-plane or out-of-plane, depending on the registry. Surprisingly, the optical transition dipole of the spin-triplet exciton is in the same order of magnitude to that of the spin-singlet exciton. So the two spin species of excitons become comparably bright in heterobilayers. This is in sharp contrast to the monolayer excitons where the spin triplet can be considered as dark compared to singlet. In a large scale moiré pattern, the local-to-local variation of the interlayer atomic registry leads to spatially modulated light coupling of interlayer exciton wavepackets. We show that the light emission polarisation and propagation direction smoothly cross between in-plane and out-of-plane ones at different locals in the moiré supercell. For both the spin-triplet and spin-singlet interlayer excitons, the overall emission into the in-plane light propagation channels has comparable strength to the out-of-plane one, another sharp difference from the monolayers where the exciton emission is predominantly into the out-of-plane light propagation channel. At given (local) atomic registry, the spin-triplet and spin-singlet excitons have distinct valley polarisation selection rules, allowing the selective optical addressing of both the valley configuration and the singlet/triplet spin configuration of interlayer excitons.

The rest of the paper is organized as follows. In Sec. II we summarize the symmetry dictated polarisation selection rules in monolayer TMDs with the presence of spin-orbit coupling. Then we go to heterobilayers in Sec. III, where the interband transitions in lattice-matched heterobilayers are discussed using both the symmetry analysis and first-principles calculations. In Sec. IV, we focus on the distinct kinematic momentum and crystal momentum of interlayer excitons in heterobilayer moiré superlattices. Based on the results of Sec. III and IV, the local optical selection rules of exciton wavepackets are then discussed in Sec. V. We summarize our results in Sec. VI.

## II.   Symmetry dictated polarisation selection rules in presence of spin-orbit coupling

Before going to the heterobilayers, we first summarize the polarisation selection rules of the monolayer excitons for comparison, and explain how the spin-orbit coupling (SOC) leads to the spin-flip interband optical transition while the out-of-plane mirror symmetry restricts that such transition can only couple to photons of out-of-plane polarisation [9,23]. The optical properties of monolayer TMDs are dominated by the ground state (1s) monolayer excitons in $\pm\mathbf{K}$ valleys. Due to the momentum conservation, the direct interconversion between such an exciton and a photon is allow only when the exciton center-of-mass (COM) momentum is nearly 0 (within the light cone). The transition dipole of the bright exciton is proportional to the optical matrix element between the valence and conduction bands at $\pm\mathbf{K}$, which then characterizes its light-coupling properties.

The SOC spin-flip term $\Delta\hat{H} = \frac{\lambda}{2\hbar}(\hat{L}_+\hat{S}_- + \hat{L}_-\hat{S}_+)$ leads to a small but finite mixing between the spin-up ($\uparrow$, $S_z = \frac{1}{2}$) and spin-down ($\downarrow$, $S_z = -\frac{1}{2}$) [23]. By treating $\Delta\hat{H}$ as a first order perturbation, the Bloch eigenstate for the $n$-th band can be written as

$$|\psi_{n,\mathbf{k},S_z}\rangle \approx |n,\mathbf{k},S_z\rangle + \frac{\lambda}{2\hbar}\sum_l \frac{\langle l,\mathbf{k},\bar{S}_z|(\hat{L}_+\hat{S}_- + \hat{L}_-\hat{S}_+)|n,\mathbf{k},S_z\rangle}{E_{n,\mathbf{k},S_z} - E_{l,\mathbf{k},\bar{S}_z}}|l,\mathbf{k},\bar{S}_z\rangle,$$

where $|n,\mathbf{k},S_z\rangle \equiv |n,\mathbf{k}\rangle \otimes |S_z\rangle$ is the unperturbed Bloch state (i.e., without considering the SOC spin-flip term), with $n$ being the band index, and $\mathbf{k}$ the crystal momentum.

The monolayer Hamiltonian conserves the $2\pi/3$-rotational ($\hat{C}_3$) symmetry and the out-of-plane mirror ($\hat{\sigma}_h$) symmetry [2], see Fig. 1(a-b). The interband transitions at $\pm\mathbf{K}$ are then governed by these symmetries. Here the $\hat{C}_3$ rotation applied on a function $\psi(\mathbf{r})$ is defined as an in-plane counterclockwise $120°$-rotation of $\psi(\mathbf{r})$ around a certain rotation center, or equivalently an in-plane clockwise $120°$-rotation of the $\mathbf{xy}$ coordinate. $\psi(\mathbf{r})$ then transforms as $\hat{C}_3\psi(\mathbf{r}) = \psi(\hat{C}_3^{-1}\mathbf{r})$. The Bloch functions at the high symmetry $\pm\mathbf{K}$ satisfy

$$\hat{C}_3 \psi_{n,S_z} = e^{-i\frac{2\pi}{3}(C_3(n)+S_z)} \psi_{n,S_z}, \qquad (1)$$
$$\hat{\sigma}_h \psi_{n,S_z} = \sigma_h(n) e^{i\pi S_z} \psi_{n,S_z}.$$

Here we have written $\psi_{n,S_z} \equiv \psi_{n,\mathbf{K},S_z}$, while the transformation of $\psi_{n,-\mathbf{K},S_z}$ by the same symmetry operations can be obtained by a time reversal. $C_3(n) = 0, \pm 1$ and $\sigma_h(n) = \pm 1$ are respectively the $\hat{C}_3$ and $\hat{\sigma}_h$ quantum numbers of the unperturbed Bloch state $|n, \mathbf{K}\rangle$ [2], i.e., $\hat{C}_3 |n, \mathbf{K}\rangle = e^{-i\frac{2\pi}{3}C_3(n)} |n, \mathbf{K}\rangle$ and $\hat{\sigma}_h |n, \mathbf{K}\rangle = \sigma_h(n) |n, \mathbf{K}\rangle$.

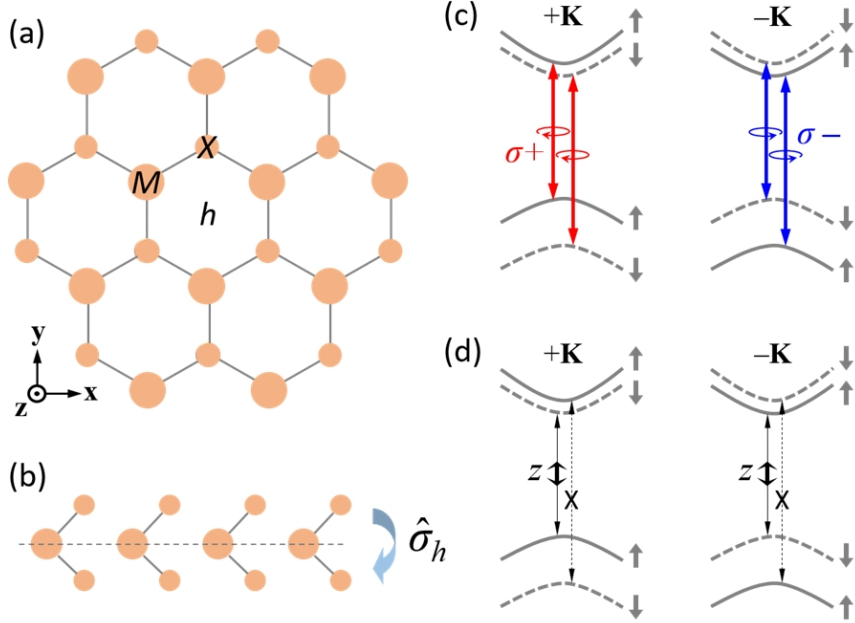

Fig. 1. (a) The top-view of a TMD monolayer, showing the $\hat{C}_3$ rotation centers $M$ (metal site), $X$ (chalcogen site), and $h$ (hexagon center). (b) The side-view of the monolayer showing the out-of-plane mirror symmetry ($\hat{\sigma}_h$). (c) The polarisation selection rules for the spin-conserved interband transitions in $\pm \mathbf{K}$ valleys. (d) The polarisation selection rules for the spin-flip interband transitions.

|  | $v-3$ | $v-2$ | $v-1$ | $v$ | $c$ | $c+1$ | $c+2$ | $c+3$ |
|---|---|---|---|---|---|---|---|---|
| $C_3(h)$ | $-1$ | $+1$ | $-1$ | $0$ | $+1$ | $0$ | $-1$ | $-1$ |
| $C_3(X)$ | $0$ | $-1$ | $0$ | $+1$ | $-1$ | $+1$ | $0$ | $0$ |
| $C_3(M)$ | $+1$ | $0$ | $+1$ | $-1$ | $0$ | $-1$ | $+1$ | $+1$ |
| $\sigma_h$ | $+1$ | $-1$ | $-1$ | $+1$ | $+1$ | $-1$ | $+1$ | $-1$ |

Table I. The $\hat{C}_3$ quantum numbers of the unperturbed Bloch state $|n, \mathbf{K}\rangle$ for rotation centers $h$, $X$ and $M$ for the band indices $n = v - 3$ to $c + 3$, as well as their $\hat{\sigma}_h$ quantum numbers. The values are taken from Ref. [2].

Due to the monolayer inversion symmetry breaking, the $\hat{C}_3$ quantum number at $\pm \mathbf{K}$ varies with the convention of the crystalline orientation. For our convention as shown in Fig. 1(a) and with $\mathbf{K} \equiv \frac{4\pi}{3a}(1,0)$, the corresponding $\hat{C}_3$ and $\hat{\sigma}_h$ quantum numbers of different bands at $\mathbf{K}$ are summarized in Table I. Here $a$ is the monolayer lattice constant. If the monolayer has an in-plane 180°-rotation compared to Fig. 1(a) (e.g., those in Ref. [21,22]), then the resulted $\hat{C}_3$ quantum numbers at $\mathbf{K} \equiv \frac{4\pi}{3a}(1,0)$ are related to Table I through a factor of $-1$. Meanwhile $\hat{\sigma}_h$ is not changed by the 180°-rotation.

The interband transitions at $\mathbf{K}$ are characterized by the optical matrix element $\langle \psi_{m,S_z'} | \hat{\mathbf{p}} | \psi_{n,S_z} \rangle$, where $\hat{\mathbf{p}} = \hat{p}_+ \mathbf{e}_+^* + \hat{p}_- \mathbf{e}_-^* + \hat{p}_z \mathbf{z}$ is the momentum operator. $\mathbf{e}_\pm \equiv (\mathbf{x} \pm i\mathbf{y})/\sqrt{2}$ with $\mathbf{x}$, $\mathbf{y}$ and $\mathbf{z}$ the three Cartesian unit vectors. Note that the $\mathbf{e}_\pm$ and $\mathbf{z}$ components of the optical matrix element couple to the in-plane $\sigma\pm$ circular and out-of-plane (z) linear polarisation of a photon, respectively. Using $\hat{C}_3 \hat{p}_\pm \hat{C}_3^{-1} = e^{\mp i\frac{2\pi}{3}} \hat{p}_\pm$, $\hat{C}_3 \hat{p}_z \hat{C}_3^{-1} = \hat{p}_z$, $\hat{\sigma}_h \hat{p}_\pm \hat{\sigma}_h^{-1} = \hat{p}_\pm$ and $\hat{\sigma}_h \hat{p}_z \hat{\sigma}_h^{-1} = -\hat{p}_z$, we obtain

$$\langle \psi_{m,S_z'} | \hat{p}_\pm | \psi_{n,S_z} \rangle = \langle \psi_{m,S_z'} | \hat{C}_3^{-1} \hat{C}_3 \hat{p}_\pm \hat{C}_3^{-1} \hat{C}_3 | \psi_{n,S_z} \rangle = e^{i\frac{2(C_3(m)-C_3(n)+S_z'-S_z \mp 1)\pi}{3}} \langle \psi_{m,S_z'} | \hat{p}_\pm | \psi_{n,S_z} \rangle$$
$$= \langle \psi_{m,S_z'} | \hat{\sigma}_h^{-1} \hat{\sigma}_h \hat{p}_\pm \hat{\sigma}_h^{-1} \hat{\sigma}_h | \psi_{n,S_z} \rangle = \sigma_h(n)\sigma_h(m) e^{i\pi(S_z-S_z')} \langle \psi_{m,S_z'} | \hat{p}_\pm | \psi_{n,S_z} \rangle,$$

$$\langle \psi_{m,S_z'} | \hat{p}_z | \psi_{n,S_z} \rangle = \langle \psi_{m,S_z'} | \hat{C}_3^{-1} \hat{C}_3 \hat{p}_z \hat{C}_3^{-1} \hat{C}_3 | \psi_{n,S_z} \rangle = e^{i\frac{2(C_3(m)-C_3(n)+S_z'-S_z)\pi}{3}} \langle \psi_{m,S_z'} | \hat{p}_z | \psi_{n,S_z} \rangle$$
$$= \langle \psi_{m,S_z'} | \hat{\sigma}_h^{-1} \hat{\sigma}_h \hat{p}_z \hat{\sigma}_h^{-1} \hat{\sigma}_h | \psi_{n,S_z} \rangle = -\sigma_h(n)\sigma_h(m) e^{i\pi(S_z-S_z')} \langle \psi_{m,S_z'} | \hat{p}_z | \psi_{n,S_z} \rangle.$$

The above equations imply that, the interband transition between $\psi_{n,S_z}$ and $\psi_{m,S_z'}$ couples to a photon with polarisation

$$\sigma\pm, \text{when:} \begin{cases} C_3(m) - C_3(n) + S_z' - S_z = 3N \pm 1, \\ \text{and} \\ \sigma_h(n)\sigma_h(m) e^{i\pi(S_z-S_z')} = 1, \end{cases} \quad (2)$$

$$z, \text{when:} \begin{cases} C_3(m) - C_3(n) + S_z' - S_z = 3N, \\ \text{and} \\ \sigma_h(n)\sigma_h(m) e^{i\pi(S_z-S_z')} = -1. \end{cases}$$

Here $N = 0, \pm 1, \pm 2, \cdots$ is a integer.

Below we shall focus on the conduction ($c$) and valence ($v$) bands, which have the properties $C_3(c) - C_3(v) = 3N + 1$ and $\sigma_h(c) = \sigma_h(v) = +1$. We can see that due to the conservation of $\sigma_h$ quantum number, spin-conserved transitions (with $S_z = S_z'$) can only couple to in-plane polarised photons while spin-flip transitions (with $S_z \neq S_z'$) can only couple to z-polarised photons [23]. Furthermore, the $\hat{C}_3$ symmetry at $\mathbf{K}$ ($-\mathbf{K}$) requires that the spin-conserved valence-to-conduction transition couples to a $\sigma+$ ($\sigma-$) photon, which is the well known valley optical selection rule in monolayer TMDs [9,10] (see Fig. 1(c)). For the spin-flip process, the criterion of Eq. (2) can only be satisfied with $S_z' = -\frac{1}{2}$ and $S_z = \frac{1}{2}$, which means that the transition $\psi_{v,\uparrow} \leftrightarrow \psi_{c,\downarrow}$ can be induced by a z-polarised photon [23]. On the other hand $\psi_{v,\downarrow} \leftrightarrow \psi_{c,\uparrow}$ is completely dark (cannot be induced by any single photon process, see Fig. 1(d)).

The spin-flip interband transitions can be understood as being mediated by the SOC spin-flip term. A perturbative treatment gives

$$\langle \psi_{c\downarrow} | \hat{\mathbf{p}} | \psi_{v\uparrow} \rangle \approx \mathbf{z} \frac{\lambda}{2\hbar} \sum_n \left( \frac{\langle c,\downarrow | \hat{p}_z | n,\downarrow \rangle \langle n,\downarrow | \hat{L}_+ \hat{S}_- | v,\uparrow \rangle}{E_{v,\uparrow} - E_{n,\downarrow}} + \frac{\langle c,\downarrow | \hat{L}_+ \hat{S}_- | n,\uparrow \rangle \langle n,\uparrow | \hat{p}_z | v,\uparrow \rangle}{E_{c,\downarrow} - E_{n,\uparrow}} \right)$$
$$\approx \mathbf{z} \frac{\lambda}{2\hbar} \left( \frac{\langle c,\downarrow | \hat{p}_z | v-2,\downarrow \rangle \langle v-2,\downarrow | \hat{L}_+ \hat{S}_- | v,\uparrow \rangle}{E_{v,\uparrow} - E_{v-2,\downarrow}} + \frac{\langle c,\downarrow | \hat{L}_+ \hat{S}_- | c+1,\uparrow \rangle \langle c+1,\uparrow | \hat{p}_z | v,\uparrow \rangle}{E_{c,\downarrow} - E_{c+1,\uparrow}} \right).$$

Note that for the bands from $v-3$ to $c+3$ as given in Table I, only the $n = v-2$ and $c+1$ intermediate bands contribute to the above spin-flip momentum matrix element. Due to the weak SOC strength $\lambda$ compared to the bandgaps $|E_{v,\uparrow} - E_{v-2,\downarrow}|$ and $|E_{c,\downarrow} - E_{c+1,\uparrow}|$, the resulted $|\langle \psi_{c,\downarrow}|\hat{p}_z|\psi_{v,\uparrow}\rangle|$ value is much smaller than the spin-conserved ones $p_0 \equiv |\langle \psi_{c,\uparrow}|\hat{p}_+|\psi_{v,\uparrow}\rangle| \approx |\langle \psi_{c,\downarrow}|\hat{p}_+|\psi_{v,\downarrow}\rangle|$. Our first-principles calculations show that $|\langle \psi_{c,\downarrow}|\hat{p}_z|\psi_{v,\uparrow}\rangle|/p_0 \approx 0.05$ for WSe$_2$ and 0.02 for MoSe$_2$, in agreement with the previous result [23].

### III. Interband transitions in lattice-matched heterobilayers

MoSe$_2$ and WSe$_2$ (MoS$_2$ and WS$_2$) are very close in lattice constants (~ 0.1%). This allows the realization of lattice-matched heterobilayer with moderate strain, for example, by CVD growth as reported in Ref. [32]. In this section we shall analyze the interband transitions for lattice-matched heterobilayers, in which the two layers have the same lattice constant and $0°$ (R-type) or $180°$ (H-type) relative rotation. The heterobilayer atomic registry is then characterized by an interlayer translation $\mathbf{r}_0$, defined as a real-space in-plane vector pointing from a metal atom in the hole layer to a nearby metal atom in the electron layer [36]. Fig. 2(a-b) illustrates the interlayer translation vector $\mathbf{r}_0$.

A heterobilayer doesn't have the $\hat{\sigma}_h$ symmetry, thus the conservation of $\sigma_h$ quantum number is no longer a constraint. However, for certain $\mathbf{r}_0$ values the $\hat{C}_3$ rotation centers of the two layers overlap, the corresponding configurations are denoted as $R_h^h$, $R_h^X$, $R_h^M$, $H_h^X$, $H_h^h$, and $H_h^M$, as shown in Fig. 2(a-b). Here $R_h^\mu$ ($H_h^\mu$) denotes the R-type (H-type) configuration with the electron-layer $\mu$ sites horizontally overlap with the hole-layer $h$ sites. For these registries the interband transition at $\pm \mathbf{K}$ are again governed by the $\hat{C}_3$ quantum number. Using the same analysis in previous section, one can show that the optical transition between $\psi_{v,S_z}$ and $\psi_{c',S_z'}$ can be induced by a photon with polarisation

$$\sigma\pm, \text{when: } C_3(c') - C_3(v) + S_z' - S_z = 3N \pm 1,$$
$$z, \text{when: } C_3(c') - C_3(v) + S_z' - S_z = 3N. \qquad (3)$$

We use the convention that quantities in the electron (hole) layer are marked with (without) the prime. Here the quantum numbers $C_3(c')$ and $C_3(v)$ should be taken around a common rotation center of the two layers. Taking $R_h^\mu$ as an example, if we set the rotation center in the hole layer as $h$, then the corresponding rotation center in the electron layer is $\mu$. Different heterobilayer atomic registries (characterized by $\mu = h, X, M$) then have different $C_3(c') - C_3(v)$ values, leading to different polarisation selection rules [36]. The resulted selection rules for valence-to-conduction transitions are summarized in Table II below.

|  | $R_h^h$ | $R_h^X$ | $R_h^M$ | $H_h^X$ | $H_h^h$ | $H_h^M$ |
|---|---|---|---|---|---|---|
| $\psi_{v,\uparrow} \leftrightarrow \psi_{c',\uparrow}$ | $\sigma+$ | $\sigma-$ | $z$ | $\sigma+$ | $\sigma-$ | $z$ |
| $\psi_{v,\downarrow} \leftrightarrow \psi_{c',\downarrow}$ | $\sigma+$ | $\sigma-$ | $z$ | $\sigma+$ | $\sigma-$ | $z$ |
| $\psi_{v,\uparrow} \leftrightarrow \psi_{c',\downarrow}$ | $z$ | $\sigma+$ | $\sigma-$ | $z$ | $\sigma+$ | $\sigma-$ |
| $\psi_{v,\downarrow} \leftrightarrow \psi_{c',\uparrow}$ | $\sigma-$ | $z$ | $\sigma+$ | $\sigma-$ | $z$ | $\sigma+$ |

Table II. The polarisation selection rules for optical transitions between the **K**-point valence band $\psi_{v,S_z}$ and conduction band $\psi_{c',S_z'}$ in $R_h^h$, $R_h^X$, $R_h^M$, $H_h^X$, $H_h^h$, and $H_h^M$ heterobilayer configurations. The band edge transitions are the $\psi_{v,\uparrow} \leftrightarrow \psi_{c',\uparrow}$ and $\psi_{v,\uparrow} \leftrightarrow \psi_{c',\downarrow}$, while the two transitions from $\psi_{v,\downarrow}$ have much higher energy.

For a quantitative analysis, the heterobilayer can be viewed as two decoupled monolayers plus a perturbative interlayer coupling term $\hat{T}$. We consider the spin-conserved or spin-flip interband transition from a valence band edge $\psi_{v,\uparrow}$ at **K** to the corresponding conduction state $\psi_{c',S_z'}$. The spin-conserved optical matrix element $\mathbf{p}_{cv,\uparrow\uparrow} \equiv \langle \psi_{c',\uparrow}|\hat{\mathbf{p}}|\psi_{v,\uparrow}\rangle$ corresponds to a first-order process mediated by the interlayer coupling $\hat{T}$:

$$\mathbf{p}_{cv,\uparrow\uparrow} \approx \sum_n \frac{\langle c',\uparrow|\hat{T}|n,\uparrow\rangle\langle n,\uparrow|\hat{\mathbf{p}}|v,\uparrow\rangle}{E_{c',\uparrow}-E_{n,\uparrow}} + \sum_{n'} \frac{\langle c',\uparrow|\hat{\mathbf{p}}|n',\uparrow\rangle\langle n',\uparrow|\hat{T}|v,\uparrow\rangle}{E_{v,\uparrow}-E_{n',\uparrow}}.$$

$$\approx \mathbf{e}_+^* \left( \frac{\langle c',\uparrow|\hat{T}|c,\uparrow\rangle\langle c,\uparrow|\hat{p}_+|v,\uparrow\rangle}{E_{c',\uparrow}-E_{c,\uparrow}} + \frac{\langle c',\uparrow|\hat{p}_+|v',\uparrow\rangle\langle v',\uparrow|\hat{T}|v,\uparrow\rangle}{E_{v,\uparrow}-E_{v',\uparrow}}\right)$$

$$+\mathbf{e}_-^* \left( \frac{\langle c',\uparrow|\hat{T}|c+2,\uparrow\rangle\langle c+2,\uparrow|\hat{p}_-|v,\uparrow\rangle}{E_{c',\uparrow}-E_{c+2,\uparrow}} + \frac{\langle c',\uparrow|\hat{p}_-|c'+2,\uparrow\rangle\langle c'+2,\uparrow|\hat{T}|v,\uparrow\rangle}{E_{v,\uparrow}-E_{c'+2,\uparrow}}\right.$$

$$\left. + \frac{\langle c',\uparrow|\hat{T}|v-3,\uparrow\rangle\langle v-3,\uparrow|\hat{p}_-|v,\uparrow\rangle}{E_{c',\uparrow}-E_{v-3,\uparrow}} + \frac{\langle c',\uparrow|\hat{p}_-|v'-3,\uparrow\rangle\langle v'-3,\uparrow|\hat{T}|v,\uparrow\rangle}{E_{v,\uparrow}-E_{v'-3,\uparrow}}\right)$$

$$+\mathbf{z}\left( \frac{\langle c',\uparrow|\hat{T}|c+1,\uparrow\rangle\langle c+1,\uparrow|\hat{p}_z|v,\uparrow\rangle}{E_{c',\uparrow}-E_{c+1,\uparrow}} + \frac{\langle c',\uparrow|\hat{p}_z|v'-2,\uparrow\rangle\langle v'-2,\uparrow|\hat{T}|v,\uparrow\rangle}{E_{v,\uparrow}-E_{v'-2,\uparrow}}\right).$$

In the above second step we have again restricted the band index summation to $\{v-3,\cdots,c+3\}$.

On the other hand the spin-flip matrix element $\mathbf{p}_{cv,\downarrow\uparrow} \equiv \langle \psi_{c',\downarrow}|\hat{\mathbf{p}}|\psi_{v,\uparrow}\rangle$ corresponds to a second-order process mediated by both the SOC spin-flip term $\frac{\lambda}{2\hbar}(\hat{L}_+\hat{S}_-+\hat{L}_-\hat{S}_+)$ and the interlayer coupling $\hat{T}$:

$$\mathbf{p}_{cv,\downarrow\uparrow} \approx \frac{\lambda}{2\hbar}\sum_{lm} \frac{\langle c',\downarrow|\hat{T}|l,\downarrow\rangle\langle l,\downarrow|\hat{L}_+\hat{S}_-|m,\uparrow\rangle\langle m,\uparrow|\hat{\mathbf{p}}|v,\uparrow\rangle}{(E_{c',\downarrow}-E_{m,\uparrow})(E_{c',\downarrow}-E_{l,\downarrow})}$$

$$+\frac{\lambda}{2\hbar}\sum_{l'm}\frac{\langle c',\downarrow|\hat{L}_+\hat{S}_-|l',\uparrow\rangle\langle l',\uparrow|\hat{T}|m,\uparrow\rangle\langle m,\uparrow|\hat{\mathbf{p}}|v,\uparrow\rangle}{(E_{c',\downarrow}-E_{m,\uparrow})(E_{c',\downarrow}-E_{l',\uparrow})}$$

$$+\frac{\lambda}{2\hbar}\sum_{ml}\frac{\langle c',\downarrow|\hat{T}|m,\downarrow\rangle\langle m,\downarrow|\hat{\mathbf{p}}|l,\downarrow\rangle\langle l,\downarrow|\hat{L}_+\hat{S}_-|v,\uparrow\rangle}{(E_{v,\uparrow}-E_{l,\downarrow})(E_{c',\downarrow}-E_{m,\downarrow})}$$

$$+\frac{\lambda}{2\hbar}\sum_{m'l'}\frac{\langle c',\downarrow|\hat{L}_+\hat{S}_-|m',\uparrow\rangle\langle m',\uparrow|\hat{\mathbf{p}}|l',\uparrow\rangle\langle l',\uparrow|\hat{T}|v,\uparrow\rangle}{(E_{c',\downarrow}-E_{m',\uparrow})(E_{v,\uparrow}-E_{l',\uparrow})}$$

$$+\frac{\lambda}{2\hbar}\sum_{l'm'}\frac{\langle c',\downarrow|\hat{\mathbf{p}}|m',\downarrow\rangle\langle m',\downarrow|\hat{L}_+\hat{S}_-|l',\uparrow\rangle\langle l',\uparrow|\hat{T}|v,\uparrow\rangle}{(E_{v,\uparrow}-E_{m',\downarrow})(E_{v,\uparrow}-E_{l',\uparrow})}$$

$$+\frac{\lambda}{2\hbar}\sum_{lm'}\frac{\langle c',\downarrow|\hat{\mathbf{p}}|m',\downarrow\rangle\langle m',\downarrow|\hat{T}|l,\downarrow\rangle\langle l,\downarrow|\hat{L}_+\hat{S}_-|v,\uparrow\rangle}{(E_{v,\uparrow}-E_{m',\downarrow})(E_{v,\uparrow}-E_{l,\downarrow})}.$$

Our previous works have shown that [38,39], under the two-center approximation and keeping only the leading Fourier components, the interlayer hopping depends on the interlayer translation $\mathbf{r}_0$ through $\langle l',S_z|\hat{T}|n,S_z\rangle \propto f_\mu(\mathbf{r}_0)$, where $\mu = \mathrm{mod}(C_3(n)-C_3(l'),3) = 0,\pm 1$ and

$$f_0(\mathbf{r}_0) \equiv e^{i\mathbf{K}\cdot\mathbf{r}_0}\left(e^{-i\mathbf{K}\cdot\mathbf{r}_0}+e^{-i\hat{C}_3\mathbf{K}\cdot\mathbf{r}_0}+e^{-i\hat{C}_3^2\mathbf{K}\cdot\mathbf{r}_0}\right)/3,$$

$$f_\pm(\mathbf{r}_0) \equiv e^{i\mathbf{K}\cdot\mathbf{r}_0}\left(e^{-i\mathbf{K}\cdot\mathbf{r}_0}+e^{-i(\hat{C}_3\mathbf{K}\cdot\mathbf{r}_0\pm\frac{2\pi}{3})}+e^{-i(\hat{C}_3^2\mathbf{K}\cdot\mathbf{r}_0\pm\frac{4\pi}{3})}\right)/3,$$

Thus the interlayer exciton optical matrix elements can have all the $\sigma+$, $\sigma-$ and $z$ components in general [38]

$$\mathbf{p}_{cv,\uparrow\uparrow}(\mathbf{r}_0) = p_{+,\uparrow\uparrow}f_0(\mathbf{r}_0)\mathbf{e}_+^* + p_{-,\uparrow\uparrow}f_+(\mathbf{r}_0)\mathbf{e}_-^* + p_{z,\uparrow\uparrow}f_-(\mathbf{r}_0)\mathbf{z}, \qquad (4)$$

$$\mathbf{p}_{cv,\downarrow\uparrow}(\mathbf{r}_0) = p_{z,\downarrow\uparrow}f_0(\mathbf{r}_0)\mathbf{z} + p_{+,\downarrow\uparrow}f_+(\mathbf{r}_0)\mathbf{e}_+^* + p_{-,\downarrow\uparrow}f_-(\mathbf{r}_0)\mathbf{e}_-^*.$$

Here $p_\mu$ are parameters which decay exponentially with the increase of interlayer distance $d$. Note that in a lattice-matched heterobilayer the interlayer distance $d$, determined by the van der Waals interaction between the layers, depends sensitively on the interlayer translation $\mathbf{r}_0$ through $d = d_0 + \Delta d_1|f_0(\mathbf{r}_0)|^2 + \Delta d_2|f_+(\mathbf{r}_0)|^2$, where the three constants $d_0$, $\Delta d_1$ and $\Delta d_2$ have been previously obtained [36]. The parameter $p_\mu$ then also depend on $\mathbf{r}_0$ through $p_\mu(d) \approx p_\mu(d_0) + \left.\frac{\partial p_\mu}{\partial d}\right|_{d=d_0}(\Delta d_1|f_0(\mathbf{r}_0)|^2 + \Delta d_2|f_+(\mathbf{r}_0)|^2)$. The $\mathbf{r}_0$-dependences of the optical matrix elements then have the forms

$$\mathbf{e}_+ \cdot \mathbf{p}_{cv,\uparrow\uparrow} \approx \left[p_{+,\uparrow\uparrow}(d_0) + \left.\frac{\partial p_{+,\uparrow\uparrow}}{\partial d}\right|_{d=d_0}(\Delta d_1|f_0(\mathbf{r}_0)|^2 + \Delta d_2|f_+(\mathbf{r}_0)|^2)\right]f_0(\mathbf{r}_0),$$

$$\mathbf{e}_- \cdot \mathbf{p}_{cv,\uparrow\uparrow} \approx \left[p_{-,\uparrow\uparrow}(d_0) + \left.\frac{\partial p_{-,\uparrow\uparrow}}{\partial d}\right|_{d=d_0}(\Delta d_1|f_0(\mathbf{r}_0)|^2 + \Delta d_2|f_+(\mathbf{r}_0)|^2)\right]f_+(\mathbf{r}_0),$$

$$\mathbf{z} \cdot \mathbf{p}_{cv,\uparrow\uparrow} \approx \left[p_{z,\uparrow\uparrow}(d_0) + \left.\frac{\partial p_{z,\uparrow\uparrow}}{\partial d}\right|_{d=d_0}(\Delta d_1|f_0(\mathbf{r}_0)|^2 + \Delta d_2|f_+(\mathbf{r}_0)|^2)\right]f_-(\mathbf{r}_0), \quad (5)$$

$$\mathbf{e}_+ \cdot \mathbf{p}_{cv,\downarrow\uparrow} \approx \left[p_{+,\downarrow\uparrow}(d_0) + \left.\frac{\partial p_{+,\downarrow\uparrow}}{\partial d}\right|_{d=d_0}(\Delta d_1|f_0(\mathbf{r}_0)|^2 + \Delta d_2|f_+(\mathbf{r}_0)|^2)\right]f_+(\mathbf{r}_0),$$

$$\mathbf{e}_- \cdot \mathbf{p}_{cv,\downarrow\uparrow} \approx \left[p_{-,\downarrow\uparrow}(d_0) + \left.\frac{\partial p_{-,\downarrow\uparrow}}{\partial d}\right|_{d=d_0}(\Delta d_1|f_0(\mathbf{r}_0)|^2 + \Delta d_2|f_+(\mathbf{r}_0)|^2)\right]f_-(\mathbf{r}_0),$$

$$\mathbf{z} \cdot \mathbf{p}_{cv,\downarrow\uparrow} \approx \left[p_{z,\downarrow\uparrow}(d_0) + \left.\frac{\partial p_{z,\downarrow\uparrow}}{\partial d}\right|_{d=d_0}(\Delta d_1|f_0(\mathbf{r}_0)|^2 + \Delta d_2|f_+(\mathbf{r}_0)|^2)\right]f_0(\mathbf{r}_0).$$

We have performed first-principles calculations for the **K**-point interband momentum matrix element of various lattice-matched heterobilayers (the calculation details is the same as in [36]). The $\sigma+$ ($\sigma-$) circularly polarised components $|\mathbf{e}_+ \cdot \mathbf{p}_{cv,\uparrow\uparrow}|/p_0$ and $|\mathbf{e}_+ \cdot \mathbf{p}_{cv,\downarrow\uparrow}|/p_0$ ($|\mathbf{e}_- \cdot \mathbf{p}_{cv,\uparrow\uparrow}|/p_0$ and $|\mathbf{e}_- \cdot \mathbf{p}_{cv,\downarrow\uparrow}|/p_0$) are shown as the red (blue) symbols in Fig. 2(c-d) at various interlayer registry $\mathbf{r}_0$, while the out-of-plane polarised components $|\mathbf{z} \cdot \mathbf{p}_{cv,\uparrow\uparrow}|/p_0$ and $|\mathbf{z} \cdot \mathbf{p}_{cv,\downarrow\uparrow}|/p_0$ are shown as the black symbols in Fig. 2(c-d). We can see that, using $p_\mu(d_0)$ and $\left.\frac{\partial p_\mu}{\partial d}\right|_{d=d_0}$ as two fitting parameters our perturbative forms (Eq. (5)) fit well the first-principles results, see the curves in Fig. 2(c-d). In lattice-matched heterobilayers, the observed interlayer exciton PL polarisation is found to be consistent with the above results [32]. For homobilayers under a large out-of-plane electric field which drives the band alignment into type-II, the interlayer exciton has the same qualitative behavior as the heterobilayer in terms of the interlayer-translation dependence of the optical matrix elements.

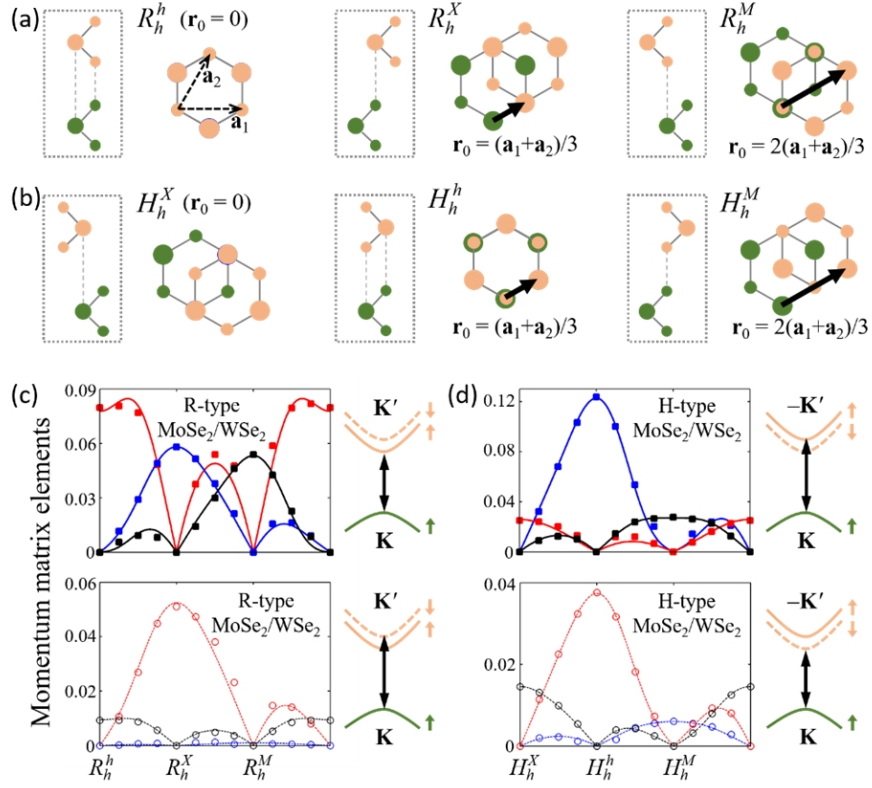

Fig. 2. (a) The three R-type $\hat{C}_3$-symmetric bilayer registries $R_h^h$, $R_h^X$ and $R_h^M$. Both the side- and top-view are shown. The dashed arrows show the monolayer unit lattice vectors. The solid arrows correspond to the interlayer translation vector $\mathbf{r}_0$. (b) The three H-type $\hat{C}_3$-symmetric bilayer registries $H_h^X$, $H_h^h$, and $H_h^M$. (c) The momentum matrix elements $\mathbf{p}_{cv,\uparrow\uparrow}$ and $\mathbf{p}_{cv,\downarrow\uparrow}$ for various R-type MoSe$_2$/WSe$_2$ heterobilayers. The red color corresponds to $|\mathbf{e}_+ \cdot \mathbf{p}_{cv}|/p_0$, blue corresponds to $|\mathbf{e}_- \cdot \mathbf{p}_{cv}|/p_0$, while black corresponds to $|\mathbf{z} \cdot \mathbf{p}_{cv}|/p_0$. $p_0$ is the interband momentum matrix element in monolayer MoSe$_2$. The symbols are obtained from first-principles calculations, while the curves are our fits using Eq. (5). (d) The momentum matrix elements for H-type MoSe$_2$/WSe$_2$ heterobilayers.

For a spin-singlet (spin-triplet) interlayer exciton, the corresponding transition dipole is proportional to the spin-conserved (spin-flip) optical matrix elements $\mathbf{p}_{cv,\uparrow\uparrow}$ ($\mathbf{p}_{cv,\downarrow\uparrow}$). The optical selection rules then depend both on the valley and interlayer atomic registry. For certain heterobilayer configurations, the spin-singlet $z$-polarised transition dipole or spin-triplet in-plane transition dipole, which are both forbidden in the monolayer, can be comparable to the maximum transition dipole strength. Both the spin-singlet and spin-triplet can then contribute significantly to the interlayer exciton PL. Moreover, in contrast to the monolayer PL that is predominantly in the out-of-plane propagation direction (as $z$-polarised transition dipole is extremely weak there, c.f. section II), significant PL in the in-plane propagation direction can also be expected in heterobilayers.

In experiments, interlayer excitons are usually generated through laser-excitation of the monolayer excitons followed by the interlayer charge transfer. It is shown that the interlayer charge transfer process largely conserves the electron/hole spin [40], thus we expect that the spin-singlet interlayer excitons are generated at the initial time, which then dominate the short-time behavior of the PL emission under a low temperature. Note that the spin-singlet and spin-triplet interlayer excitons have a small energy difference due to the conduction band spin splitting. Whether the ground state exciton is in a spin-singlet or spin-triplet configuration depends on the materials and stacking pattern. For example, in a near R-type MoSe₂/WSe₂ heterobilayer, the ground state is a spin-singlet (Fig. 2(c)). While in a near H-type MoSe₂/WSe₂ the ground state is a spin-triplet (Fig. 2(d)). In the latter case, the long-time behavior of PL can be determined by the relaxation dynamics between the spin-singlet and spin-triplet interlayer excitons. The fact that spin-triplet excitons have comparable brightness with the spin-singlet ones in general can significantly affect the overall PL property.

## IV. Heterobilayers with moiré pattern: kinematic momentum and crystal momentum of interlayer excitons

In reality, most heterobilayers host the moiré patterns due to the inevitable lattice mismatch and twist. Here we generalize the above finding in lattice-matched heterobilayers to the moiré superlattice. We denote the electron (hole) layer lattice constant as $a'$ ($a$), and $\delta\theta$ the small deviation of the twist angle to $N\pi/3$. The optically active heterobilayers are those with $\delta = |a - a'|/a' \ll 1$ and $\delta\theta \ll 1$, whose optical properties are dominated by the interlayer excitons. For such heterobilayers, $\pm \mathbf{K}'$ in the electron layer are displaced from $\pm \mathbf{K}$ in the hole layer and long-period moiré superlattice patterns form. Writing $\mathbf{b}'_{1,2}$ ($\mathbf{b}_{1,2}$) as the electron (hole) layer primitive reciprocal lattice vectors, $\mathbf{b}'_1 - \mathbf{b}_1$ and $\mathbf{b}'_2 - \mathbf{b}_2$ with $|\mathbf{b}'_{1,2} - \mathbf{b}_{1,2}| \ll b'_1, b_1$ are the primitive reciprocal lattice vectors of the heterobilayer moiré (see Fig. 3(a)). The moiré period is then $A = 4\pi/\sqrt{3}|\mathbf{b}'_1 - \mathbf{b}_1| \approx a/\sqrt{\delta^2 + \delta\theta^2}$.

Consider an interlayer exciton formed by a near-band-edge electron at $\tau'\mathbf{K}' + \mathbf{k}'$ in the Bloch form $\psi_{\tau'\mathbf{K}'+\mathbf{k}',c'}(\mathbf{r}) = e^{i(\tau'\mathbf{K}'+\mathbf{k}')\cdot\mathbf{r}} u_{\tau'\mathbf{K}'+\mathbf{k}',c'}(\mathbf{r})$, and a near-band-edge hole at $\tau\mathbf{K} - \mathbf{k}$ in the form $\psi^*_{\tau\mathbf{K}-\mathbf{k},v}(\mathbf{r}) = e^{-i(\tau\mathbf{K}-\mathbf{k})\cdot\mathbf{r}} u^*_{\tau\mathbf{K}-\mathbf{k},v}(\mathbf{r})$. Here $u_{\mathbf{k},c}$ and $u_{\mathbf{k}',v}$ are the periodic parts. The electron-hole Coulomb interaction conserves the momentum sum $\mathbf{Q} \equiv \mathbf{k} + \mathbf{k}'$ [38]. Thus it is convenient to describe the interlayer exciton in the basis of Coulomb interaction (momentum) eigenstates $X_{\tau'\tau,\mathbf{Q}}$, characterized by the COM kinematic momentum $\mathbf{Q}$ and the electron-hole valley indices $(\tau', \tau)$.

Using the COM coordinate $\mathbf{R} \equiv \frac{m_e}{M_0}\mathbf{r}_e + \frac{m_h}{M_0}\mathbf{r}_h$ and the electron-hole relative coordinate $\mathbf{r}_{eh} \equiv \mathbf{r}_e - \mathbf{r}_h$, the kinematic momentum eigenstate of interlayer excitons can be written as:

$$X_{\tau'\tau,\mathbf{Q}}(\mathbf{r}_e, \mathbf{r}_h) = \sum_{\Delta\mathbf{Q}} \Phi(\Delta\mathbf{Q}) \psi_{\tau'\mathbf{K}'+\frac{m_e}{M_0}\mathbf{Q}+\Delta\mathbf{Q},c'}(\mathbf{r}_e) \psi^*_{\tau\mathbf{K}-\frac{m_h}{M_0}\mathbf{Q}+\Delta\mathbf{Q},v}(\mathbf{r}_h) \quad (6)$$
$$= e^{i(\mathbf{Q}+\tau'\mathbf{K}'-\tau\mathbf{K})\cdot\mathbf{R}} U_{\tau'\tau,\mathbf{Q}}(\mathbf{R}, \mathbf{r}_{eh}).$$

Here $M_0 \equiv m_e + m_h$ is the exciton mass, with $m_e$ ($m_h$) the electron (hole) effective mass. $\Phi(\Delta\mathbf{Q})$ describes the momentum space electron-hole relative motion. In the above second step, we write

$$U_{\tau'\tau,\mathbf{Q}}(\mathbf{R}, \mathbf{r}_{eh}) \equiv \sum_{\Delta\mathbf{Q}} e^{i\left(\Delta\mathbf{Q} + \frac{m_h}{M_0}\tau'\mathbf{K}' + \frac{m_e}{M_0}\tau\mathbf{K}\right)\cdot\mathbf{r}_{eh}} \Phi(\Delta\mathbf{Q}) \quad (7)$$

$$\times u_{\tau'\mathbf{K}'+\frac{m_e}{M_0}\mathbf{Q}+\Delta\mathbf{Q},c'}\left(\mathbf{R}+\frac{m_h}{M_0}\mathbf{r}_{eh}\right)u^*_{\tau\mathbf{K}-\frac{m_h}{M_0}\mathbf{Q}+\Delta\mathbf{Q},v}\left(\mathbf{R}-\frac{m_e}{M_0}\mathbf{r}_{eh}\right).$$

Below we show that, $U_{\tau'\tau,\mathbf{Q}}(\mathbf{R},\mathbf{r}_{eh})$ is a periodic function of $\mathbf{R}$ with the periodicity of the moiré pattern, such that the COM *crystal momentum* of the interlayer exciton in the moiré superlattice can also be defined.

Using the electron (hole) layer reciprocal lattice vectors $\mathbf{G}' = N_1\mathbf{b}'_1 + N_2\mathbf{b}'_2$ ($\mathbf{G} = N_1\mathbf{b}_1 + N_2\mathbf{b}_2$), the Bloch periodic part can be expanded as $u_{\mathbf{k}',c'}(\mathbf{r}) = \sum_{\mathbf{G}'} e^{i\mathbf{G}'\cdot\mathbf{r}}C_{c'}(\mathbf{k}'+\mathbf{G}')$ ($u_{\mathbf{k},v}(\mathbf{r}) = \sum_{\mathbf{G}} e^{i\mathbf{G}\cdot\mathbf{r}}C_v(\mathbf{k}+\mathbf{G})$). Their product is then

$$u_{\mathbf{k}',c'}\left(\mathbf{R}+\frac{m_h}{M_0}\mathbf{r}_{eh}\right)u^*_{\mathbf{k},v}\left(\mathbf{R}-\frac{m_e}{M_0}\mathbf{r}_{eh}\right) = \sum_{\mathbf{G}'\mathbf{G}} e^{i(\mathbf{G}'-\mathbf{G})\cdot\mathbf{R}+i\left(\frac{m_h}{M_0}\mathbf{G}'+\frac{m_e}{M_0}\mathbf{G}\right)\cdot\mathbf{r}_{eh}}C_{c'}(\mathbf{k}'+\mathbf{G}')C^*_v(\mathbf{k}+\mathbf{G}).$$

We are interested in how the above form varies with $\mathbf{R}$ in a scale much larger than the monolayer lattice constant. In this case only those $e^{i(\mathbf{G}'-\mathbf{G})\cdot\mathbf{R}}$ with $|\mathbf{G}'-\mathbf{G}| \ll b'_1, b_1$ are important while the others with $|\mathbf{G}'-\mathbf{G}| \sim b'_1, b_1$ or larger are averaged out. Also $C_{c'}(\mathbf{k}'+\mathbf{G}')$ ($C_v(\mathbf{k}+\mathbf{G})$) with $G' \gg b'_1$ ($G \gg b_1$) correspond to fast oscillating terms in $u_{\mathbf{k}',c'}$ ($u_{\mathbf{k},v}$) with periods much smaller than $a'$ ($a$), which are negligibly small and can be dropped. The remaining slowly oscillating terms have $\mathbf{G}' - \mathbf{G} = N_1(\mathbf{b}'_1 - \mathbf{b}_1) + N_2(\mathbf{b}'_2 - \mathbf{b}_2)$ which are just the reciprocal lattice vectors of the moiré. Thus $U_{\tau'\tau,\mathbf{Q}}(\mathbf{R}+\mathbf{R}_{SL},\mathbf{r}_{eh}) = U_{\tau'\tau,\mathbf{Q}}(\mathbf{R},\mathbf{r}_{eh})$ is periodic in $\mathbf{R}$, where $\mathbf{R}_{SL}$ is a superlattice vector of the moiré.

Concerning the COM motion, the momentum eigenstate $X_{\tau'\tau,\mathbf{Q}} = e^{i(\mathbf{Q}+\tau'\mathbf{K}'-\tau\mathbf{K})\cdot\mathbf{R}}U_{\tau'\tau,\mathbf{Q}}(\mathbf{R},\mathbf{r}_{eh})$ is then in an analogous form to the Bloch function, with $U_{\tau'\tau,\mathbf{Q}}$ being its periodic part (with moiré periodicity instead) and a plane-wave envelop $e^{i(\mathbf{Q}+\tau'\mathbf{K}'-\tau\mathbf{K})\cdot\mathbf{R}}$. We can therefore have the notion of the *crystal momentum* of interlayer exciton in the moiré superlattice, which is $\mathbf{Q} + \tau'\mathbf{K}' - \tau\mathbf{K}$, plus a reciprocal lattice vector $\mathbf{G}' - \mathbf{G}$ of the moiré.

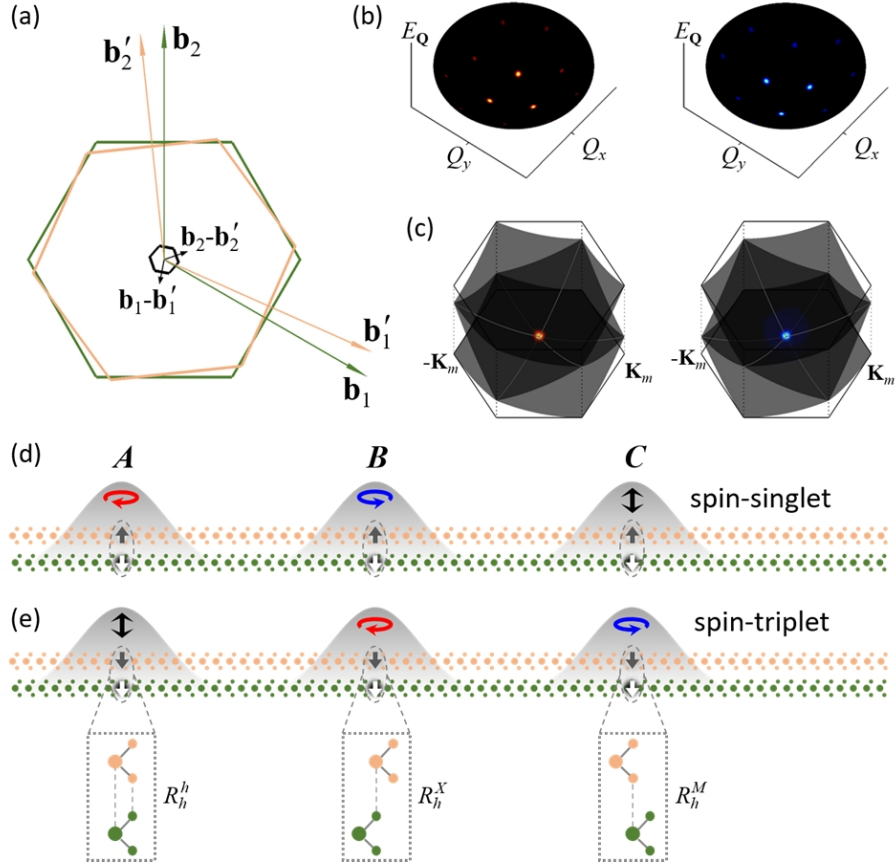

Fig. 3. (a) The orange (green) hexagon is the BZ of the electron (hole) layer, with $\mathbf{b}'_{1,2}$ ($\mathbf{b}_{1,2}$) the corresponding primitive reciprocal lattice vectors. The black hexagon is the moiré superlattice mini BZ, with $\mathbf{b}'_1 - \mathbf{b}_1$ and $\mathbf{b}'_2 - \mathbf{b}_2$ the moiré primitive reciprocal lattice vectors. (b) The interlayer exciton dispersions as functions of the kinematic momentum $\mathbf{Q}$. Left and right correspond to opposite valley indices. The light cone positions are indicated as the bright spots, where the three brightest ones are the main light cones. (c) The interlayer exciton dispersions as functions of the crystal momentum in the mini BZ of the moiré pattern, where $\pm\mathbf{K}_m$ are the mini BZ corners. Only the three lowest-energy bands are shown. Again the left and right correspond to opposite valley indices. The gray curves are the doubly degenerate nodal lines for a given valley. The light cone, located at $\mathbf{\Gamma}$, is three-fold degenerate (six if including the valley degeneracy). (d) Schematic illustration of the polarisation selection rules, for $\mathbf{K}$-valley spin-singlet interlayer exciton wavepackets centered at $A$, $B$ and $C$, respectively. (e) The polarisation selection rules for the $\mathbf{K}$-valley spin-triplet interlayer exciton wavepackets. $A$, $B$ and $C$ have the atomic registes $R_h^h$, $R_h^X$ and $R_h^M$, respectively.

Unlike the monolayer excitons and interlayer excitons in lattice-matched heterobilayers where the exciton COM crystal momentum and kinematic momentum coincide, the two momentums are different for interlayer excitons in heterobilayer moiré patterns. The kinematic momentum of $X_{\tau'\tau,\mathbf{Q}}$ is $\mathbf{Q}$ which gives its COM velocity $\langle X_{\tau'\tau,\mathbf{Q}}|\dot{\mathbf{R}}|X_{\tau'\tau,\mathbf{Q}}\rangle = \frac{\hbar}{M_0}\mathbf{Q}$ and kinetic energy $\frac{\hbar^2 Q^2}{2M_0}$ [38], while the momentum conservation in the light coupling concerns the crystal momentum in the moiré superlattice which is $\mathbf{Q} + \tau'\mathbf{K}' - \tau\mathbf{K} + \mathbf{G}' - \mathbf{G}$. Momentum conservation requires a bright interlayer exciton to have zero COM crystal momentum. Bright interlayer excitons then have finite kinematic momentums $\mathbf{Q} = \tau\mathbf{\kappa} - \tau'\mathbf{\kappa}'$ ($\tau'\mathbf{\kappa}' \equiv \tau'\mathbf{K}' + \mathbf{G}'$ and $\tau\mathbf{\kappa} \equiv \tau\mathbf{K} + \mathbf{G}$) which are termed as the light cone positions in Ref. [38]. It has been shown in Ref. [38] that, the most important light cones are those located at $\mathbf{Q}_0 = \tau\mathbf{K} - \tau'\mathbf{K}'$, $\hat{C}_3\mathbf{Q}_0$ and $\hat{C}_3^2\mathbf{Q}_0$, where the valley indices $(\tau', \tau) = (+, +)$ or $(-, -)$ for a near R-type stacking and $(\tau', \tau) = (-, +)$ or $(+, -)$ for a near H-type stacking. The other light cones have orders of magnitude weaker transition dipole strengths thus are usually ignored [38].

The interlayer exciton dispersion as a function of kinematic momentum $\mathbf{Q}$ shows a simple paraboloid, on which the light cones form an ordered array connected by the reciprocal lattice vectors of the moiré, see Fig. 3(b). Note that the $\mathbf{Q} = 0$ state is optically dark, while all the light cones have finite kinetic energies. In the crystal momentum picture, by folding the original paraboloid dispersion to the mini BZ of the moiré superlattice, the interlayer exciton dispersion forms a series of mini bands with the band width $\sim \frac{\hbar^2}{2M_0}\left(\frac{8\pi}{3A}\right)^2$ as shown in Fig. 3(c). The main light cones are located at the $\mathbf{\Gamma}$ point of the lowest-energy mini band. For each given valley, there are three doubly degenerate nodal lines which cross at $\mathbf{\Gamma}$ (see Fig. 3(c)). $\mathbf{\Gamma}$ then has a degeneracy of three (six if including the valley degeneracy). These nodal lines are unstable, whose degeneracies disappear when taking into account the moiré superlattice potential [36,37] originating from the local-to-local variation of the bandgaps in the moiré [35]. Under a general moiré superlattice potential there is no degenerate point for each given valley, and the mini bands can be very different from those in Fig. 3(c) (see for example Ref. [36]). However for certain moiré potential patterns the degeneracies at the high-symmetry points $\mathbf{\Gamma}$ and the mini BZ corners $\pm\mathbf{K}_m$ can be partly restored. As pointed out in Ref. [36], an out-of-plane electric field can tune the interlayer exciton moiré potential to a honeycomb shape, where $\mathbf{\Gamma}$ is a four-fold degenerate Dirac node while $\pm\mathbf{K}_m$ are doubly degenerate Weyl nodes.

## V. Local optical selection rules in the moiré: from momentum eigenstates to exciton wavepackets

In a moiré potential, the description of interlayer excitons can also be facilitated using the picture of wavepackets, whose COM spatial extension $w$ shall be small compared to the moiré period $A$. Excitons in different supercells are coupled through COM motion induced hopping, which leads to mini bands [36,37]. The hopping is determined by the competition between the kinetic energy $\frac{\hbar^2}{2M_0}\left(\frac{8\pi}{3A}\right)^2$ and the potential barrier. For moiré patterns with small periods, where $\frac{\hbar^2}{2M_0}\left(\frac{8\pi}{3A}\right)^2$ is comparable to the moiré potential modulation range and the hopping strength is significant, it is convenient to start with the momentum eigenstates, the coupling of which by the moiré superlatttice potential leads to the folding into mini bands as discussed in section IV (also see Ref. [36]). For large enough periods, the hopping between neighboring supercells are small and the low energy bands become nearly flat [36,37]. In this case it is more convenient to start from interlayer exciton wavepackets moving in the slowly varying moiré potential, where the reminiscent hopping between neighboring supercells also leads to mini band formations.

The moiré period $A$ can be as large as several nm to several tens nm. At any local region with a length scale small compared to $A$ but much larger than the monolayer lattice constant $a$, the corresponding atomic registry is nearly indistinguishable from a lattice-matched heterobilayer. The electronic structure and the optical properties of such a local region can then be described by those of the lattice-matched heterobilayer [35,41,42]. For a qualitative analysis, the optical properties of the interlayer exciton wavepacket should be determined by the local atomic registry within the wavepacket extension. At different places in the moiré supercell, the interlayer exciton wavepacket will have its properties resembling excitons in lattice-matched heterobilayers of different interlayer registries, which vary smoothly from local-to-local. To be specific, spin-singlet (spin-triplet) wavepackets centered at $A$, $B$ and $C$ with $R_h^h$, $R_h^X$, $R_h^M$ registries couple to $\sigma+$, $\sigma-$ and $z$ ($z$, $\sigma+$ and $\sigma-$) polarised photons, respectively, see Fig. 3(d-e).

The above physical picture is corroborated by a quantitative evaluation of the optical transition dipole of an interlayer exciton wavepacket in the moiré superlattices. A low-energy wavepacket can be constructed with the basis of momentum eigenstates [36]

$$\mathcal{X}_{\tau'\tau,\mathbf{R}_c}(\mathbf{R},\mathbf{r}_{eh}) = \sum_{\mathbf{Q}} e^{-i(\mathbf{Q}-\mathbf{Q}_0)\cdot\mathbf{R}_c} W(\mathbf{Q}) X_{\tau'\tau,\mathbf{Q}}(\mathbf{R},\mathbf{r}_{eh}). \tag{8}$$

Here $\mathbf{R}_c$ is the wavpacket center position. $W(\mathbf{Q})$ is the kinematic momentum distribution function, which is centered at $\mathbf{Q}=0$ and has a width $\sim 2/w$ inversely proportional to the real space extension $w$. As the transition dipole of $X_{\tau'\tau,\mathbf{Q}}$ is only finite at the three main light cones $\mathbf{Q}_0$, $\hat{C}_3\mathbf{Q}_0$ and $\hat{C}_3^2\mathbf{Q}_0$, the transition dipole of the wavepacket is the coherent superposition of those from $X_{\tau'\tau,\mathbf{Q}_0}$, $X_{\tau'\tau,\hat{C}_3\mathbf{Q}_0}$ and $X_{\tau'\tau,\hat{C}_3^2\mathbf{Q}_0}$ with their phase factors $1$, $e^{-i(\hat{C}_3\mathbf{Q}_0-\mathbf{Q}_0)\cdot\mathbf{R}_c}$ and $e^{-i(\hat{C}_3^2\mathbf{Q}_0-\mathbf{Q}_0)\cdot\mathbf{R}_c}$ depending on the wavepacket center [36]. Using the previously obtained $X_{\tau'\tau,\mathbf{Q}}$ transition dipole [38], we find the wavepacket transition dipole forms

$$\propto W(\mathbf{Q}_0)[p_{+,\uparrow\uparrow}f_0(\mathbf{r}_0(\mathbf{R}_c))\mathbf{e}_+^* + p_{-,\uparrow\uparrow}f_+(\mathbf{r}_0(\mathbf{R}_c))\mathbf{e}_-^* + p_{z,\uparrow\uparrow}f_-(\mathbf{r}_0(\mathbf{R}_c))\mathbf{z}], \tag{9}$$
$$\propto W(\mathbf{Q}_0)[p_{z,\downarrow\uparrow}f_0(\mathbf{r}_0(\mathbf{R}_c))\mathbf{z} + p_{+,\downarrow\uparrow}f_+(\mathbf{r}_0(\mathbf{R}_c))\mathbf{e}_+^* + p_{-,\downarrow\uparrow}f_-(\mathbf{r}_0(\mathbf{R}_c))\mathbf{e}_-^*],$$

for the spin-singlet and spin-triplet interlayer excitons, respectively. They are nearly the same as Eq. (4) for the lattice-matched heterobilayer, except that the constant interlayer translation vector $\mathbf{r}_0$ is now replaced by a position-dependent local value $\mathbf{r}_0(\mathbf{R}_c)$. Note that the wavepacket real space extension $w$ is small compared to the moiré period $A$, thus the width $2/w$ of $W(\mathbf{Q})$ is much larger than $Q_0 = \frac{4\pi}{3A}$. So $W(\mathbf{Q}_0) \approx W(0)$ is not exponentially small, and such a wavepacket is always bright.

Eq. (9) shows that the optical properties of the wavepacket is indeed determined by the local atomic registry within the wavepacket extension. Both the strength of the optical transition dipole and the valley optical selection rules are spatially modulated with the wavepacket center. Remarkably, at the $\hat{C}_3$ symmetric locals $R_h^h$, $R_h^X$, $R_h^M$, $H_h^X$, $H_h^h$, and $H_h^M$ (correspond to extrema of the moiré superlattice potential [36]), the interlayer exciton wavepackets can emit $\sigma+$, $\sigma-$ and $z$-polarised photons for a given valley configuration (Fig. 3(d-e)).

## VI. Summary and discussion

We have investigated the light-coupling properties of both the spin-singlet and spin-triplet interlayer excitons in TMD heterobilayers. For an incommensurate heterobilayer with a large scale moiré pattern, we establish a one-to-one correspondence between the local interlayer atomic registry and the optical transition dipole of the interlayer exciton wavepacket. This correspondence is found to be isomorphic to the one between the interlayer registry in lattice-matched heterobilayers and the optical transition dipole therein, suggesting the validity of using local approximations to describe the optical and electronic properties in the long period Moiré superlattices. The local-to-local variation of the atomic registry in the long period Moiré pattern then leads to spatially modulated light-coupling of interlayer excitons.

These findings can change the interpretation and design of optical experiments in heterobilayers, and leads to new geometries for optoelectronic devices based on interlayer excitons. For example, in a near R-type $MoSe_2/WSe_2$ the interlayer exciton has its energy minima of the moiré potential at **B** locations with $R_h^X$ registry [36]. Valley polarised spin-singlet and spin-triplet interlayer excitons at these locations emit circularly polarised photons with opposite polarisation as shown in Fig. 2(c). This could be the origin of the recently observed two interlayer exciton PL peaks with opposite circular polarisation [43]. Meanwhile, the spin-singlet and spin-triplet also have distinct location-dependence of the out-of-plane optical transition dipole (c.f. Fig. 2(c)). A nano-optical antenna-tip can be used to greatly ($\sim 6 \times 10^5$) enhance the PL from the out-of-plane transtion dipoles [25]. With a spatial resolution of the tip enhanced PL ($\leq 15$ nm) comparable to the moiré supercell size, the nano-optical antenna-tip should be able to distinguish the distinct out-of-plane transtion dipole strengths of the spin-singlet and spin-triplet at given locations in the moiré .

**Acknowledgments:** The work is supported by the Croucher Foundation (Croucher Innovation Award), the Research Grants Council (HKU17312916), and the University of Hong Kong (ORA). G.-B.L. was supported by NSFC with Grant No. 11304014 and the China 973 Program with Grant No. 2013CB934500.

**Appendix: The first-principles calculations**

The first-principles calculations are performed using the Vienna Ab-initio Simulation Package [44] based on plane waves and the projector-augmented wave (PAW) method [45]. The Perdew-Burke-Ernzerhof (PBE) [46] exchange-correlation functional is used for all calculations and the van der Waals interactions are considered in the DFT-D3 [47] method. The experimentally measured bulk lattice constants are respectively 3.288 Å for $MoSe_2$ [48] and 3.282 Å for $WSe_2$ [49]. Their average of 3.285 Å is used for the lattice-matched $MoSe_2/WSe_2$ heterobilayer. Keeping the in-plane positions fixed, the out-of-plane positions are relaxed for all atoms until the energy difference of successive atom configurations is less than $10^{-6}$ eV. The out-of-plane force on each atom in the relaxed structure is less than 0.003 eV/Å. The cutoff energy of the plane-wave basis is set to 350 eV and the convergence criterion for total energy is $10^{-8}$ eV. A Γ-centered **k** mesh of $15 \times 15 \times 1$ is used for both the relaxation and normal calculations. The thickness of vacuum layer is greater than 20 Å to avoid impacts from neighboring periodic images. Spin-orbit coupling is taken into account for all calculations except in structure relaxation.